# Retention Time of Peptides in Liquid Chromatography Is Well Estimated upon Deep Transfer Learning

Chunwei Ma[1,2,#], Zhiyong Zhu[3,#], Jun Ye[3], Jiarui Yang[1,4], Jianguo Pei[3], Shaohang Xu[1,2], Chang Yu[3], Fan Mo[3], Bo Wen[1,2,*], Siqi Liu[1,2,*]

1 BGI-Shenzhen, Shenzhen 518083, China.
2 China National GeneBank, BGI-Shenzhen, Shenzhen 518083, China.
3 Intel Asia-Pacific Research & Development Ltd, Shanghai 200241, China.
4 Qingdao University-BGI Joint Innovation College, Qingdao University, Qingdao 266071, China.
# Equal contribution
* Correspondence:
Dr. Siqi Liu, BGI-Shenzhen, Shenzhen, 518083, China, E-Mail: siqiliu@genomics.cn
Bo Wen, BGI-Shenzhen, Shenzhen, 518083, China, E-Mail: wenbo@genomics.cn

**Abbreviations**:
**LC**, liquid chromatography;
**RT**, retention times;
**CNN**, Convolutional Neural Network;
**ResNet**, Residual Network;
**LSTM**, Long Short-Term Memory;
**TL**, Transfer learning;
**SGD**, stochastic gradient descent;
**RMSE**, root-mean-square error;
**DDA**, data-dependent acquisition;
**DIA**, data-independent acquisition.

**Abstract**: A fully automatic prediction for peptide retention time (RT) in liquid chromatography (LC), termed as DeepRT, was developed using deep learning approach, an ensemble of Residual Network (ResNet) and Long Short-Term Memory (LSTM). In contrast to the traditional predictor based on the hand-crafted features for peptides, DeepRT learns features from raw amino acid sequences and makes relatively accurate prediction of peptide RTs with 0.987 $R^2$ for unmodified peptides. Furthermore, by virtue of transfer learning, DeepRT enables utilization of the peptides datasets generated from different LC conditions and of different modification status, resulting in the RT prediction of 0.992 $R^2$ for unmodified peptides and 0.978 $R^2$ for post-translationally modified peptides. Even though chromatographic behaviors of peptides are quite complicated, the study here demonstrated that peptide RT prediction could be largely improved by deep transfer learning. The DeepRT software is freely available at https://github.com/horsepurve/DeepRT, under Apache2 open source License.

## 1 Introduction

In liquid chromatography-coupled tandem mass spectrometry-based (LC-MS/MS-based) proteomics, peptides are separated by liquid chromatography coupled with typically reverse phase column (RPLC), prior being induced into a mass spectrometer. The difference between the beginning of a chromatographic gradient until a specific peptide elutes from the liquid chromatography (LC) and injected into mass spectrometer is referred as that peptide's retention time (RT). The RT is dependent on the physical and chemical properties of the peptide, and is specific to the given LC condition. The RT is reproducible over experiments, and is to some extent predictable.[1] The different between observed and predicted RT can hence be used to eliminate the false positive of peptide or protein identifications both in bottom-up and top-down proteomics.[2] Moreover, as fragment spectra are less unique in data-independent acquisition (DIA) mass spectrometry-based proteomics, peptide RTs can provide unexpendable information for the identification of a peptide.[1] In Selected Reaction Monitoring (SRM), the accurately predicted RTs are useful to target the interested peptides on liquid chromatogram.[3] As the peptide RT is advantageous for analysis and interpretation of experimental data, how to gain its prediction within a tolerant error range has been attracting research attentions for decades.

Most RT predictors focus on feature engineering, exploring the relationship of peptide RTs with the sophisticated and comprehensive features for peptides as many as possible. The early RT prediction model took solely amino acid composition as retention coefficients (RC).[4] Subsequent studies tended to integrate more peptide features into estimation of RTs, such as residue position,[5] peptide length or mass, hydrophobic moment[6] and topology.[7] Petritis et al. proposed the secondary structural contents (SSC) could be used for RT prediction.[6] As SSC is generally resulted from prediction algorithm, such RT evaluation is restricted by how SSC is well designed. Posttranslational modifications (PTMs) is another important aspect related with RTs. In general, however, the confident amount of modified peptides is usually limited so that their effects to RT are not easily estimated. A software termed as Elude was regarded as one of the most widely used RT predictors, which designed a set of 60 features combining both general hydrophobicity index and retention coefficients trained for the specific dataset.[3] Although Elude was demonstrated as a good predictor in a lot of testing datasets,[8] it was still criticized by lack of appropriate features representing the secondary structure of peptide.[1] Since all the hand-crafted features or rules are restricted within personal expertise or knowledge, it is generally recognized that such limitation could slow down the algorithm development for RT prediction. A question is naturally raised whether the peptide features serving for RT prediction could be generated from a machine analysis, a non-personal experience dependent approach. Žuvela et al. pointed out that the peptide features could be selected through quantitative structure-retention relationship (QSRR) model, and the selected features really benefited for RT prediction.[9] However, the QSRR selection was still partially relied on the hand-crafted features. More recently, Krokhin et al. adopted SSRCalc to incorporate the sequence and helical features for RT prediction, and observed that the prediction accuracy was impacted by the liquid chromatographic conditions, hydrophilic interaction column (HILIC) different from RPLC.[10] A peptide RT predictor that is derived from features independent from human expertise and is designed by machine learning has yet to be developed.

In recent years, deep learning has made groundbreaking achievements in the realms of image and speech recognition,[11] automatic game-playing of Go,[12] sequence specificity prediction,[13] cancer diagnosis,[14] and protein structure prediction.[15] With deep architectures, secondary structure elements of proteins, such as helix, coil and sheet, were predictable only on the basis of amino acid sequences. Hence, it is reasoned that the peptide RTs could be forecasted using deep learning as well.[1] Moreover, an important characteristic of deep learning is fine-tuning mechanism that allows transferring common characteristics across over different datasets.[16] The approach thus is expected powerful for RT prediction through variated peptide datasets, including peptide data generated from differently experimental conditions.

Herein, we proposed the DeepRT software, which utilizes the advanced deep learning frameworks for peptide RT prediction. Basically, DeepRT uses a 5-layers Residual Network (ResNet) and a 2-layers Long Short-Term Memory (LSTM), and performs end-to-end feature extraction and non-linear regression without manual intervention. DeepRT is capable of improvement of RT prediction by using multiple datasets that are produced from different sources such as organisms, LC conditions and PTMs.

## 2 Materials and Methods

### 2.1 Datasets

The two datasets provided by Moruz et al.[17] were used for evaluation of the software developed in this study, in which one was from mouse liver enriched for phosphopeptides[18] and the other one was from yeast[19]. Another dataset was obtained from Rosenberger et al.,[20] which contained the peptides generated from analysis of human proteomes. The proteomics data in the three datasets was generated in $C_{18}$ column but their acetonitrile gradients were different (see **Supporting Information Table T-1** for details). The identified peptides in the datasets were directly taken for training and testing of RT prediction algorithms. The number of unique peptides, retention times and lengths of the all peptides in three datasets were illustrated in **Table 2** and **Supporting Information Figure S-1**.

### 2.2 LSTM

The LSTM approach[21] was introduced for peptide RT prediction. LSTM is a variant of Recurrent Neural Network (RNN) with the specially designed memory cell that can capture important long-term information, and have been applied in protein study.[11,22] Generally, LSTM takes elements one by one, and gives an output after taking every single element. In this study all the LSTM outputs were discarded but only the final one after inputting all elements (amino acids) was remained, which was expected as the approximate RT of a peptide.

Formally, in order to build a general framework for RT prediction of both unmodified and unmodified peptides, we define an ordered amino acid set as $S$, in which the unmodified and modified form of an amino acid are two distinct elements in it. In addition, we introduce a pseudo amino acid denoted by *O into $S$ in order that all peptides have the same length. For example, if the maximum length of all peptides $l$ is 10, and then the peptide VGEDIAK is padded to be OOOVGEDIAK***VGEDIAK. Therefore, for unmodified peptides with 20 kinds of amino acids, in the set $S$, the number of elements $|S|$ is 21., whereasile for dataset with only one kind of modification, $|S|$ is 22.

A memory cell takes an amino acid as input at each time step, and has a cell state $c_t$ and a hidden state $h_t$, in which the time step t represents the $t^{th}$ position of amino acid at a peptide sequence. The cell state and hidden state are changed after each time step, then fed into the same memory cell itself along with the next input vector at the next step. The hidden state of a memory cell in certain LSTM layer severs as input to the cell in the next LSTM layer. Specifically, the hidden state at time step $t$ is calculated as,

$$i_t = \sigma(W_{xi}x_t + W_{hi}h_{t-1} + b_i)$$
$$f_t = \sigma(W_{xf}x_t + W_{hf}h_{t-1} + b_f)$$
$$\tilde{c}_t = \tanh(W_{xc}x_t + W_{hc}h_{t-1} + b_c)$$

$$c_t = f_t \odot c_{t-1} + i_t \odot \tilde{c}_t$$
$$o_t = \sigma(W_{xo}x_t + W_{ho}h_{t-1} + b_o)$$
$$h_t = o_t \odot \tanh(c_t)$$

where $\odot$ denotes the Hadamard product, $\sigma(\cdot)$ is sigmoid function that scale the value to the (0,1) range:

$$\sigma(x) = \frac{1}{1+e^{-x}}$$

and $i, f, o \in [0,1]^{d_h}$ are the input, forget and output gates. The weights $W_{x\cdot} \in \mathbb{R}^{d_h \times d_x}$, $W_{h\cdot} \in \mathbb{R}^{d_h \times d_h}$ and bias $b_i, b_f, b_c, b_o \in \mathbb{R}^{d_h}$ are the model parameters where $d_h$ is the number of hidden neurons in a memory cell and note that $x_t \in \mathbb{R}^{d_x}$ is the embedding vector of dimension $d_x$ for the amino acid at time step $t$. The cell output $h_t \in [-1,1]^{d_h}$ and cell state $c_t \in \mathbb{R}^{d_h}$ are all vectors.

Suppose that there are $n_{lstm}$ LSTM layers in the network, then the output of the last memory cell at the final time step is $h_l^{(n_{lstm})}$ where $l$ is the maximum length of all peptides. In order to explore the nonlinear relationship of features in $h_l^{(n_{lstm})}$, we append a fully-connected layer with $d_{fc}$ neurons, behind the last memory cell. Thus the output of such layer is:

$$h_{fc} = \sigma\left(W_{fc}h_l^{(n_{lstm})} + b_{fc}\right)$$

where $W_{fc} \in \mathbb{R}^{d_{fc} \times d_h}$ and bias $b_{fc} \in \mathbb{R}^{d_{fc}}$ are parameters.

Finally, the output of the overall network i.e. retention time is computed as,

$$RT_{lstm} = W_{lstm}h_{fc} + b_{lstm}$$

where $W_{lstm}$ and $b_{lstm}$ are parameters for the linear regression. The structure of LSTM for RT prediction is illustrated in the upper left part of **Figure 1.** The LSTM network is trained using stochastic gradient descent (SGD).

### 2.3 ResNet

The peptide RT is estimated with ResNet[23] in this study as well, an improved version of convolutional neural network (CNN). ResNet introduces the identity shortcut connections and allows several stacked layers to fit a residual mapping of $\mathcal{F}(X) = \mathcal{H}(X) - X$, where $\mathcal{H}(X)$ is the original and desired mapping.

One-hot encoding of amino acid is set as the input to ResNet, in which each amino acid is represented as a $|S|$-dimensional vector. For a dataset, if the maximum length of all peptides is $l$, then a peptide is represented by a $|S| \times l$ matrix $X$ with components $X_{ij}$ satisfying,

$$X_{ij} = \begin{cases} 1, & p_j = s_i \\ 0, & otherwise \end{cases}$$

where $p_j$ indicates the $j$th amino acid of the input peptide and $s_i$ denotes the $i$th amino acid in $S$.

Before utilizing shortcut connections, a convolutional layer with $n_{conv}$ filters is taken to detect the low-level patterns. The output of this convolutional layer is a $(l + 1 - m_{conv}) \times n_{conv}$ matrix $X^{(conv)}$, where $m_{conv}$ is the filter size. Element $X_{i,k}^{(conv)}$ is a filtered feature of filter $k$ aligned to position $i$ of the peptide. The parameters are stored in an $n_{conv} \times m_{conv} \times |S|$ matrix $M$, where element $M_{k,j,g}$ is the coefficient of filter

$k$ at filter position $j$ and amino acid $g$. Mathematically, the expression $X^{(conv)} = conv_M(X)$ computes a discrete cross-correlation between the input peptide matrix $X$ and each filter $M_k$, where

$$X^{(conv)}_{i,k} = ReLU\left(\sum_{j=1}^{m_{conv}} \sum_{g=1}^{|S|} X_{i+j,g} M_{k,j,g}\right)$$

for all $1 \leq i \leq l + 1 - m_{conv}$ and $1 \leq k \leq n_{conv}$, while $ReLU$ represents the rectified linear function:

$$ReLU(x) = \begin{cases} x, & x \geq 0 \\ 0, & x < 0 \end{cases}$$

And $X^{(conv)}$ severs as input to a residual unit. The residual unit consists of two convolutional layers with filter size of $m_{res}$ and its output is computed as

$$X^{(res_1)} = conv\left(conv(X^{(conv)})\right) + X^{(conv)}$$

and $X^{(res_1)}$ will in turn to be the input to the following residual units. Suppose there are $n_{res}$ residual units in all, then the output of the last residual unit $X^{(res_{n_{res}})}$ is the highest level features for RT prediction and the retention time is computed as

$$RT_{res} = W_{res} X^{(res_{n_{res}})} + b_{res}$$

where $W_{res}$ and $b_{res}$ are parameters for the linear regression. The structure of ResNet for RT prediction is illustrated in the upper right part of **Figure 1**. The deep residual network is trained using stochastic gradient descent (SGD).

## 2.4 Data augmentation, Ensemble and transfer learning for peptide retention time prediction

Besides, several efforts to further improve the performance of RT prediction were made. As LSTM makes prediction only after the final time step, the impact of the last amino acid of a peptide to network should be weighted carefully. A technique named data augmentation[24] is used to address this issue. For every peptide in training set, all the amino acid sequences are reversed and the reversed sequences are added into training set. In testing set, the model will make prediction to the original and reversed peptides respectively and gives two values, $\widehat{RT}_{lstm}$, and $\widehat{RT}_{lstm-rev}$. As for RetNet, it generally learns shift-invariant features so that the treatment of reversed peptides is not required.

An ensemble learning approach is utilized to integrate the different prediction values and to achieve better performance than the constituent predictor alone.[25] The ensemble of LSTM and ResNet is shown in bottom of **Figure 1.** For a peptide in training dataset, the trained model predicts three distinct values, $RT_{lstm}$, $RT_{lstm-rev}$ and $RT_{res}$. To have an ensemble of the three predicted values and the observed peptide RT, a Multiple Linear Regression (MLR) model is applied,

$$RT = A_{lstm} RT_{lstm} + A_{lstm-rev} RT_{lstm-rev} + A_{res} RT_{res}$$

At the testing stage, the learned parameters for MLR are used to calculate the final prediction of each peptide:

$$\widehat{RT} = A_{lstm} \widehat{RT}_{lstm} + A_{lstm-rev} \widehat{RT}_{lstm-rev} + A_{res} \widehat{RT}_{res}$$

Previous study used model selection and calibration to transfer the learned models to new datasets.[3] In their method, the trained models were directly applied to the training set of the new data, and the Spearman's rank correlation between the predicted and observed RT in training set was taken to evaluate the prediction models. After the best model was determined, a regression model was trained. to calibrate the previous trained model. However, with Regarding the flexibility of deep learning to transfer parameters from different datasets, transfer learning[26] was used onthe pre-trained models. First, a model was trained on a large dataset, and its parameters were used to initialize the new deep network model for a new dataset, then the new model was trained in the new dataset as normal to update the new parameters.

The DeepRT software includes LSTM, ResNet and ensemble pipeline, which are publicly available MXNet library.[27]

# 3 Results

### 3.1 Optimization of deep network architectures

In deep learning, parameter tuning is necessary to acquire the appropriate hyper-parameters and to build the network architectures with a good performance for computational prediction. In this study, the small mouse dataset with 3413 peptides was taken to optimize the network architectures in either LSTM or ResNet. The dataset was split resulting in 90% of the peptides as the training set, while the other as testing one, and the splitting process was performed 3 times. Greedy algorithm was applied to the 3 training-testing pairs to find the optimized hyper-parameters, that is, first determining the best value of one hyper-parameter, which is then fixed during the optimization of the rest hyper-parameters. Based on the hyper-parameters selected, we extended them in other larger datasets.

In LSTM, the parameter tuning resulted in the optimized hyper-parameters, like the dimension of embedding vector $d_x$, the hidden neurons in the memory cells $d_h$, the number of LSTM layers $n_{lstm}$, number of neurons in the final sigmoid layer $d_{fc}$ and dropout. The parameter-tuning process of LSTM is shown in **Supporting Information Table S-1** and the optimized parameters are listed in **Table 1**. As indicated by the tables, LSTM model with 2 layers could achieve better performance such as Pearson's correlation and Δt$_{95\%}$ (discussed below), and run economically because additional layers could not significantly improve the prediction performance.

In ResNet, there were 5 hyper-parameters to be optimized, such as number of filters $n_{conv}$ in every convolutional layers, filter size for the first convolutional layer $m_{conv}$ and the following residual units $m_{res}$, number of residual units $n_{res}$, and dropout ratio. The parameter-tuning process of ResNet is exhibited in **Supporting Information Table S-2** and the final chosen parameters are listed in **Table 1**. These results demonstrated that ResNet with 2 residual units was able to achieve better prediction performance and to run efficiently.

The outputs of LSTM and ResNet were further undergone the ensemble treatment with MLR. As presented in **Supporting Information Table S-7 and S-8**, the prediction performance after the treatment was improved in the two datasets.

### 3.2 Improvement of RT Prediction by Deep Learning

The evaluation of the RT prediction was carried out in three software, ELUDE and GPTime which are well accepted in the field at present, and DeepRT developed in this study. With consideration of peptide contents, peptide sources, and peptide modification status, three datasets were selected, mouse with 3413 peptides, yeast with 14361 peptides and human with 146587 peptides. Generally, a software was applied to an individual dataset for RT prediction, during which the dataset was randomly divided to two parts, training and testing with random seeds up to 10.

The software performance for RT prediction was judged by four metrics, in which Pearson's correlation coefficient indicates how close of the RT values between predicted and observed ones, the squared correlation $R^2$, root-mean-square error (RMSE) represents the average prediction error, and $\Delta t_{95\%}$ means the minimal time window containing the deviations between observed and predicted RT for 95% of the peptides. All the metrics estimated by the three software were depicted in **Figure 2,** and the details for each run against mouse and yeast datasets were summarized in **Supporting Information Table S-3~S-8**, respectively. Clearly, in the two datasets, the RT prediction performance of DeepRT exhibited overall advantages to the other two software (**Figure 2**), while Elude and GPTime showed similar performance due to the features shared by them. Specifically, the average Pearson's correlation gained from DeepRT was 0.982 (0.966 $R^2$) for mouse dataset and 0.994 (0.987 $R^2$) for yeast dataset, whereas, that from Elude and GPTime were 0.977 and 0.976 for mouse, and 0.981 and 0.981 for yeast. The values of $\Delta t_{95\%}$ based on DeepRT were significantly reduced by 22.5% (16.6 min to 12.8 min) for mouse dataset and 48.7% (48.1 min to 24.7 min) for yeast dataset, as compared with the values elicited from Elude. The comparison results demonstrated that the algorithm upon deep learning could provide more accurate RT prediction than that generated from personal experience. Besides, we also took DeepRT to predict RT for human dataset, during which 90% of the 146587 peptides were randomly picked out with 3 different seeds. On the human dataset, we had to exclude Elude and GPTime from comparison because of their prohibitively long running times on such a large dataset. The result shows that the average Pearson's correlation is 0.996 (0.993 $R^2$) with $\Delta t_{95\%}$ being 14.64 min. The predicted and observed RT and distribution of the prediction errors are shown in **Figure 4** and the results for all three random experiments are listed in **Supporting Information Table S-9** and the predicted retention times for each peptide are listed in **Supporting Information Table S-10**.

### 3.3 Improvement of RT Prediction by Transfer Learning

Previous study utilized the training data of new dataset for calibration of the trained models, and the calibrated and selected model was used for prediction of the testing set. The question is that such RT prediction approach may not well fit into other datasets due to data generation at different conditions such as LC and modification status, even after the model calibration. Differently, with the flexibility of deep learning to transfer parameters across different datasets, we cast the calibration problem into transfer learning task: we trained a model on a large human dataset and then fine-tuned it on the mouse and yeast datasets.

Instead of predicting RT only using single dataset described above, a model was pre-trained on a non-target dataset, and was fine-tuned on the target dataset to gain the final RT prediction. For instance, for the RT prediction for mouse dataset, the pre-trained human RT model was generated by DeepRT, and it was further fine-tuned on the mouse dataset. Similarly, for the RT prediction of yeast peptides, the pre-trained treatment was done on human dataset, then re-trained process was implemented in yeast dataset. As shown in Figure 2, the Pearson's correlation increased from 0.989 (0.978 $R^2$) for mouse (modified) dataset and 0.996 (0.992 $R^2$) for yeast (unmodified) dataset. Meanwhile, the $\Delta t_{95\%}$ value decreased to 8.85 min and 18.02 min, 46.7% and 62.5% less than Elude, for the two datasets respectively. Taken the evidences described above, these results demonstrated that DeepRT was capable of utilization of the trained model for RT prediction, and further improve the accuracy.

Finally, we tested whether this scheme could be generalized to different chromatographic column types. Traditionally, RT prediction under a certain HPLC column was conducted using data solely from such column type, due to the different separation mechanism of different columns. Notwithstanding, the separations of peptides still share some commonness under different column types, which, we expected, could be utilized by transfer learning. We thus fine-tuned the trained human model on an in-house rice datasets generated under strong cation exchange chromatography (SCX) with 5574 peptides in it. The result shown that, compared with using only SCX data, the accuracy of prediction was improved from 0.923 to 0.943 Pearson's correlation, with $\Delta t_{95\%}$ decreased from 9.70 min to 8.32 min, by incorporating the pre-trained RP-HPLC model.

**3.4 Running efficiency of DeepRT**

For sake of DeepRT running efficiency, we evaluated its performance in two aspects, how much time and how big dataset were required to get a better prediction. In mouse dataset, the training process averagely cost 1.5 and 1.1 hours for Elude and GPTime, whereas only 0.4 hours for DeepRT. In yeast dataset, the training time was taken to 47.2 and 30.7 hours on average for Elude and GPTime, whereas was dramatically shrunk to 1.29 hours by DeepRT. In some cases, limited dataset may affect the RT prediction. DeepRT seems to estimate a better RT prediction within limited peptides. For instance, DeepRT gave comparable prediction (0.982 Pearson correlation and 46.01 $\Delta t_{95\%}$ on average) with Elude and GPTime, which was only based on one ninth training data available (1437 peptides).

As the amount of training data is important for the precise prediction of RT, we examined DeepRT's performance on training sets with gradually increased peptide numbers from 0.1 to 8 times of the same testing set (1437 peptides). As shown in **Figure 2**, the performance of DeepRT increases dramatically with adding training peptides from 143 to 1437 and keeps on increasing while adding more peptides.

**4 Discussion**

As peptide RTs are governed by myriad factors, accurate prediction to the values is difficultly relied on personal expertise. We therefore developed a software for RT prediction, termed as DeepRT, which utilizes deep learning technique without any required features. This software takes the raw peptide sequences as input and extracts features automatically, and learns sophisticated relationship of the sequences and RTs through multiple and stacked layers (**Supporting Information Table S-15**). DeepRT enables a relatively precise prediction for RTs and substantially outperforms the current predictors such as Elude and GPTime.

With a traditional approach of RT prediction, a trained model for one dataset cannot be directly applied to another dataset based upon the same chromatographic resin, because there are the irregular non-linear shifts of peptide RTs due to different LC conditions[28]. Petritis et al. pointed out that the RT prediction model required 345914 training peptides to get fine accuracy, which made the algorithm difficult to retrain for other chromatographic conditions[1]. Although this problem would be alleviated by the alignment and calibration treatment towards the RTs in different datasets, the accuracy of RT prediction would be suffered significantly. Herein, DeepRT demonstrated it's capable of dealing with multiple datasets to facilitate the precise RT prediction. As revealed in Figure 3, DeepRT didn't only work with three datasets for RT prediction, but also its prediction indeed got further improvement during expansion of dataset sizes. Besides, DeepRT could run the prediction to the datasets in which the peptides are generated from different post-translational modifications. Moruz et al. proposed to incorporate unmodified peptide into training set for the RT prediction of modified peptides[29]. Instead of integration, DeepRT can directly combine the dataset containing the peptides with/without post-translational modification, thus benefit both prediction accuracy and speed. The prediction really helps to understand of how amino acid residue modifications influence peptide RTs.

In Supplementary Figures S3-S5, we plotted the predicted RTs by DeepRT versus the observed RTs for all the peptides in testing sets, mouse dataset (**Supporting Information Figure S-3**) and yeast dataset (**Supporting Information Figure S-4**). The predicted RTs for a few of peptides obviously deviate far from the corresponding observed RTs on the two figures, implying that DeepRT prediction does not work well for all the testing peptides. How does DeepRT mis-predicted those peptide RTs? We did the prediction with other predictors to the same peptide dataset, such as Elude and GPTime. Unexpectedly, the three algorithms gave a similar result as DeepRT, in which the predicted RTs for the same peptides even only a few of them appeared the deviated trends. As all the four algorithms reach to the comparable prediction, this means that the predicted deviation was not caused by the predictors, but was likely to come from the peptides themselves. If the prediction errors was plotted against the error frequencies (**Supporting Information Figure S-5**), the distribution curve of DeepRT displayed a good symmetry centered to zero as compared with that generated from the other three software, however, a small error shoulders were found all the distribution curves, suggesting a consistent error of RT prediction towards the same peptide dataset. Furthermore, we also found that the less accuracy the RT prediction is, the longer peptides are, probably due to limited long peptides in training set. Considering peptide identification most derived from MS-based approach, the identification errors are generally recognized due to the false in detection signals or annotation software. We therefor inquire to if the prediction deviation for RT is resulted from the false positive identification to the peptides.

In Table 3, we demonstrated DeepRT achieved high efficiency in program running. How does it perform so well? Generally, given the fixed feature number, support vector machine (SVM) and Gaussian process regression should have time complexities as $O(n^2)$ and $O(n^3)$, where $n$ is the number of training data. On the other hand, DeepRT utilizes a technique so called stochastic gradient descent (SGD), in which all the update network parameters typically concern only several dozen samples with mini-batch mode, thus time complexity of SGD is $O(n)$. The learning curve of DeepRT is illustrated in **Supporting Information Figure S-2**. A proper function of time complexity is a main reason leading to the running efficiency of DeepRT on relatively large dataset.

The DeepRT software is freely available at https://github.com/horsepurve/DeepRT, under Apache2 open source License.

Funding: This study was supported by the International Science & Technology Cooperation Program of China (2014DFB30020).

The authors have declared no conflicts of interest.

# Figures and Tables

**Table 1**. Model structure of DeepRT. The meanings of each hyper-parameters are described in the main text.

| | $d_x$ | $d_h$ | $n_{lstm}$ | $d_{fc}$ | dropout ratio |
|---|---|---|---|---|---|
| LSTM | 20 | 128 | 2 | 256 | 0.2 |
| | $n_{conv}$ | $m_{conv}$ | $m_{res}$ | $n_{res}$ | dropout ratio |
| ResNet | 20 | 5 | 11 | 2 | 0 |

**Table 2**. List of the datasets used in this study. The performance of mouse and yeast datasets were obtained using DeepRT with transfer learning (TL) on 10 random experiments, while the performance of human was calculated using DeepRT on 3 random experiments.

| Datasets | Modification | RT (min) | unique peptides | training peptides | testing peptides | Pearson | $R^2$ | $\triangle t_{95\%}$ (min) |
|---|---|---|---|---|---|---|---|---|
| mouse | modified | 0~109 | 3413 | 3071 | 342 | 0.989 | 0.978 | 8.85 |
| yeast | unmodified | 0~263 | 14361 | 12924 | 1437 | 0.996 | 0.992 | 18.02 |
| human | unmodified | -60~183* | 146587 | 131928 | 14659 | 0.996 | 0.993 | 14.65 |

*The retention times of human dataset were normalized using iRT Kit.

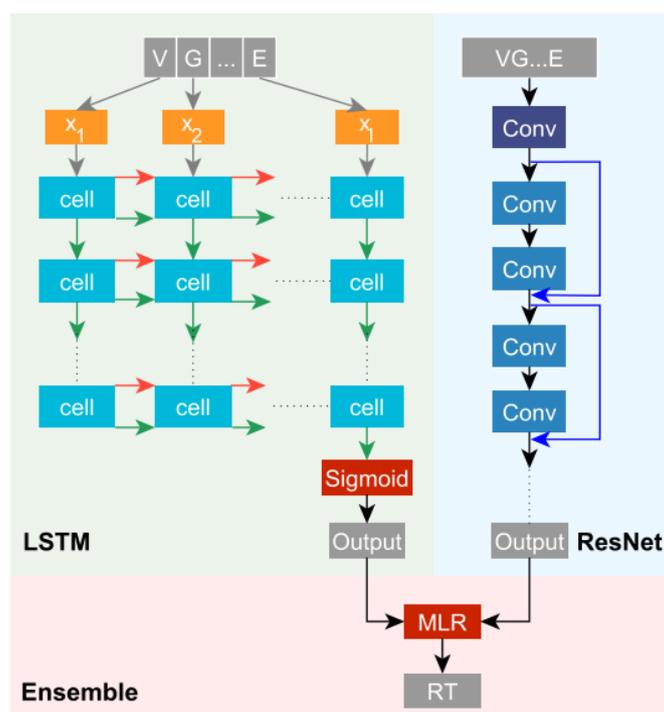

**Figure 1**. Overview of the DeepRT pipeline. For a peptide input, its RT is predicted by LSTM or ResNet in parallel, and the predicted values are ensembled by MLR to generate the final RT prediction.

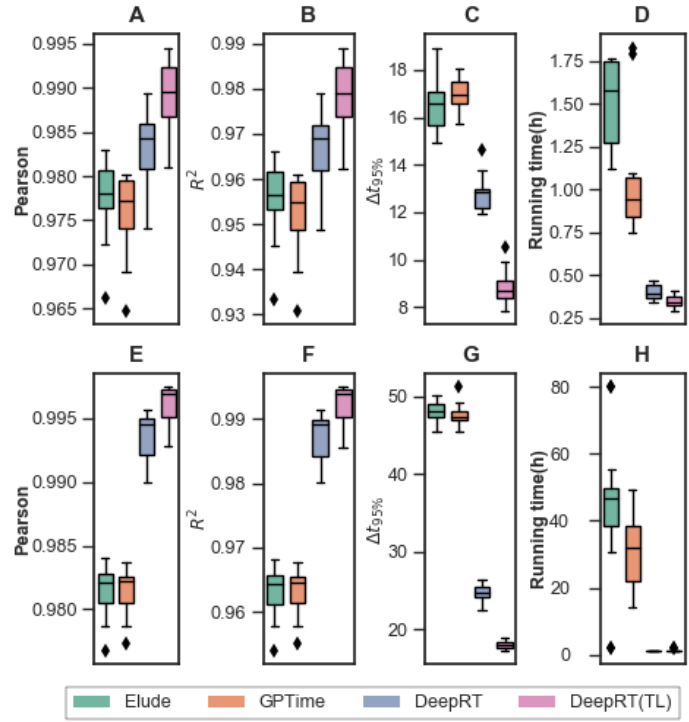

**Figure 2.** Performance comparison of three RT predictors, Elude, GPTime and DeepRT. All the predictors were employed to two datasets, mouse dataset (upper panel) and yeast dataset (lower panel). Each boxplot was plotted using 10 random runs.

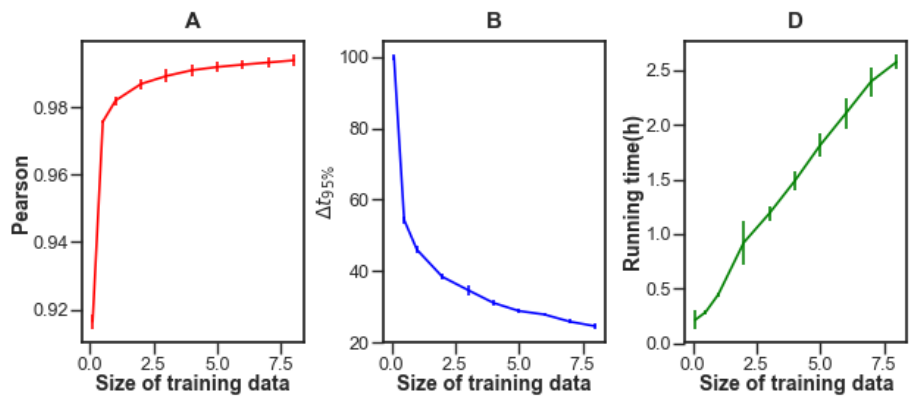

**Figure 3.** The performance of DeepRT on 1437 testing peptides of yeast dataset as a function of the size of training data. The performance was measures as Pearson correlation (A) and $\triangle t_{95\%}$ (B); the running times were reported too (C). The horizontal axis indicates the ratio of training data to testing data. The error bars show the standard deviation of 3 random experiments.

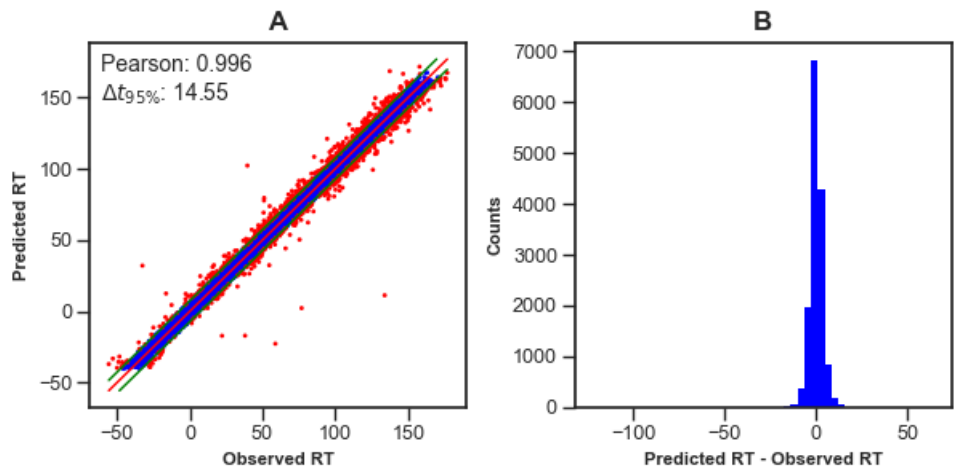

**Figure 4**. DeepRT's performance on human dataset in terms of the relationship of predicted RT and observed RT of 14659 testing peptides (A) and the distribution of the prediction error (B) of human dataset. The red points in A indicate the peptides whose prediction deviation fall outside the $\triangle t_{95\%}$ window.